\documentclass[10pt,aps,prd,onecolumn,nofootinbib,showpacs,showkeys,superscriptaddress]{revtex4-2}

\usepackage{psfrag}
\usepackage{mathrsfs}
\usepackage{amssymb, bm}
\usepackage{amsmath, amsthm}
\usepackage{epstopdf}
\usepackage[breaklinks=true]{hyperref}
\usepackage{enumerate}
\usepackage[ansinew]{inputenc}
\usepackage{longtable}
\usepackage{subfigure}
\usepackage{color}
\usepackage{mathrsfs}
\usepackage{graphicx}
\usepackage[table]{xcolor}
\usepackage{color}
\usepackage{bm,natbib,url,textcase}
\usepackage{xurl}
\usepackage[title]{appendix}

\usepackage[english]{babel}
\usepackage[T1]{fontenc}
\usepackage{comment}
\numberwithin{equation}{section}
\usepackage{geometry}
\allowdisplaybreaks

\hypersetup{
 colorlinks=true, 
 linkcolor=blue, 
 citecolor=blue, 
 urlcolor=blue 
}

\usepackage{tikz}

\definecolor{purple}{rgb}{1,0,1}
\definecolor{lime}{HTML}{a6CE39} 

\newcommand{\orcidicon}{%
	\begin{tikzpicture}
		\draw[lime, fill=lime] (0,0) 
		circle [radius=0.15] 
		node[white] {{\fontfamily{qag}\selectfont \tiny ID}};
		\draw[white, fill=white] (-0.0625,0.095) 
		circle [radius=0.007];
	\end{tikzpicture}	\hspace{-2mm}
}

\newcommand\orcidValerio{{\href{https://orcid.org/0000-0002-2601-1870}{\orcidicon}}}
\newcommand\orcidAndrea{{\href{https://orcid.org/0000-0003-0329-2726}{\orcidicon}}}
\newcommand\orcidLuca{{\href{https://orcid.org/0009-0006-3167-4990}{\orcidicon}}}

\def\T{{\cal T}}
\def\K{{\cal K}}

\newcommand{\be}{\begin{equation}}
\newcommand{\ee}{\end{equation}}

\newcommand{\red}{\color{red}}
\def \d {{\mathrm{d}}}
\newcommand{\affbish}{\affiliation{Department of Physics \& Astronomy, Bishop's University, 2600 College Street, Sherbrooke, Qu\'ebec, Canada J1M~1Z7 }}
\newcommand{\affilAMtwo}{\affiliation{Alma Mater Research Center on Applied Mathematics (AM2) Via Saragozza 8, 40123 Bologna, Italy}}

\begin{document}

\title{Towards a causal  effective thermodynamics of scalar-tensor gravity}

\author{Narayan Banerjee}
\email[]{narayan@iiserkol.ac.in}
\affiliation{Department of Physical Sciences, Indian  Institute of Science 
Education and Research Kolkata, Mohanpur 741 246, West Bengal, India}

\author{Andrea Giusti\orcidAndrea}
\email[]{andrea.giusti9@unibo.it}
\affiliation{Department of Physics and Astronomy "A. Righi", University of Bologna,  via Irnerio 46, 40126 Bologna, Italy}\
\affilAMtwo
\affiliation{INFN, Sezione di Bologna, IS FLAG viale Berti Pichat 6/2, 40127  Bologna, Italy}

\author{Valerio Faraoni\orcidValerio}
\email[]{vfaraoni@ubishops.ca}
\affbish

\author{Luca Gallerani\orcidLuca}{\normalsize {\large }}
\email[]{luca.gallerani6@unibo.it}
\affiliation{Department of Mathematics, University of Bologna, Piazza di porta San Donato 5, 40126 Bologna, Italy}
\affilAMtwo

\author{Alain Maltais-Gosselin}
\email[]{amaltais26@ubishops.ca}
\affbish
{\LARGE {\Large }}

\begin{abstract} 
The thermal analogy between the effective fluid of 
scalar-tensor gravity and Eckart's irreversible thermodynamics is extended 
to the causal Israel-Stewart model, adopting the minimal 
{\em ansatz} of promoting the heat 
flux density to a timelike vector. This choice yields analytically 
manageable constitutive equations, allowing for the first consistent 
decoupling of the effective temperature $\cal{T}$ and the effective 
thermal conductivity $\cal{K}$ of scalar-tensor gravity. Crucially, this 
new framework preserves the interpretation of general relativity as the  
equilibrium state approached via a dynamical relaxation process in the 
vanishing-${\cal KT}$ limit. This new causal formalism is applied to 
cosmology. 

\end{abstract} \maketitle

\section{Introduction}
\label{sec:1}
\setcounter{equation}{0}

There are strong indications that General Relativity (GR), might not be 
the ultimate theory of gravity. Indeed, despite its success and its 
powerful predictions, it still suffers from the problem of spacetime 
singularities. Furthermore, at cosmological scales, the standard $\Lambda$CDM model 
relies heavily on a dark sector \cite{Amendola:2015ksp}, which is affected 
by worrisome tensions \cite{Riess:2019qba,DiValentino:2021izs}. To address these issues, one 
option is that GR could be the low-energy limit of another more general 
theory. The simplest GR extension is scalar-tensor gravity, which 
introduces a scalar field $\phi$ in the Einstein-Hilbert action, nonminimally 
coupled with the gravitational sector 
\cite{Brans:1961sx,Bergmann:1968ve,Nordtvedt:1968qs, 
Wagoner:1970vr,Nordtvedt:1970uv}. 
The progenitor of scalar-tensor  
gravity is Brans-Dicke gravity \cite{Brans:1961sx}. Their idea was that a 
proper gravitational theory should incorporate Mach's principle, which is 
implemented by the relation $\phi \simeq 1/G_{\text{eff}}$, i.e., the 
scalar field as the inverse of the effective gravitational coupling 
strength. This new ingredient entails a dynamical Newton constant $G$, as 
first proposed by Dirac \cite{Dirac:1938mt}. In parallel, one of the most 
popular alternatives to GR is \textit{$f({\cal R})$} gravity, where 
${\cal R}$ is the 
Ricci scalar and $f$ a non linear function 
\cite{Sotiriou:2008rp,DeFelice:2010aj,Nojiri:2010wj}. Notably 
$f({\cal R})$ gravity turns out to be a subclass of scalar-tensor gravity, which 
makes this 
class of theories even more appealing. \par Scalar-tensor gravity has  
recently been investigated from a thermal perspective, originating an 
analogy{\huge {\Huge }} called \textit{first order thermodynamics of 
scalar-tensor gravity} 
\cite{Faraoni:2021lfc,Faraoni:2021jri,Giusti:2021sku,Giardino:2022sdv} 
(see \cite{Giardino:2023ygc} for an exhaustive review). Within this 
framework, the authors formalized a map between the constitutive equations 
of Eckart's irreversible thermodynamics \cite{Eckart:1940te} and the 
dissipative components of the effective energy-momentum tensor resulting 
from the modified field equations. Remarkably, this procedure provides an 
expression for the quantity ${\cal KT}$, where ${\cal K}$ is an effective 
thermal conductivity and ${\cal T}$ the so-called effective 
\textit{temperature of gravity} whose value determines the relaxation of 
scalar-tensor gravity to (or departure from) GR. This thermal analogy 
interprets  scalar-tensor 
gravity as an out-of-equilibrium state that can dynamically relax toward 
GR in the limit ${\cal KT} \to 0$. This feature provides an elegant 
interpretation of the \textit{attractor-to-GR mechanism} 
\cite{Faraoni:2025alq,Giusti:2026ymb}, a topic extensively analyzed by 
Damour and Nordtvedt in \cite{Damour:1992kf,Damour:1993id}, generating 
a large literature and debate. \par The main limitation of the 
thermal view 
of scalar-tensor gravity is shared with Eckart's irreversible 
thermodynamics, which is known to be non-causal, despite having been 
successfully and widely used for many years. The goal of this work is to 
extend the thermal view of scalar-tensor gravity to the Israel-Stewart 
model \cite{Israel:1979wp}, which entails hyperbolic, causal constitutive 
relations. In the remainder of this section we briefly review 
scalar-tensor gravity, while the thermal analogy is recalled in 
Sec.~\ref{sec:2}, followed by its causal extension in Sec.~\ref{sec:3}. We 
then apply this new formalism to Bianchi~I and 
Friedmann-Lema\^itre-Robertson-Walker (FLRW) universes in 
Secs.~\ref{sec:4} and~\ref{sec:5} and conclude with an analysis of exact 
solutions in Sec.~\ref{sec:6}.

\par
The action of scalar-tensor gravity reads\footnote{We shall follow the notation of Ref.~\cite{Waldbook} in which $G=c=1$, with $G$ and $c$ Newton's constant and speed of light respectively}:
\be
S_\mathrm{ST} =  \int d^4x \, \frac{\sqrt{-g}}{16\pi}  \left[ \phi {\cal 
R} 
-\frac{\omega(\phi )}{\phi} 
\, \nabla^c\phi \nabla_c\phi -V(\phi) \right] +S^\mathrm{(m)} \,,  
\label{STaction}
\ee
where ${\cal R}\equiv { {\cal R}^c}_c $ is the Ricci scalar, the 
Brans-Dicke scalar $\phi>0$ is 
approximately the inverse of the 
effective gravitational 
coupling, $\omega(\phi)$ is the ``Brans-Dicke coupling'', $V(\phi)$ is a 
potential,  $g$ is the determinant of the metric and $S^\mathrm{(m)}=\int 
d^4x \sqrt{-g} \, 
{\cal  L}^\mathrm{(m)} $ is the  
matter action. The  field equations 
\cite{Brans:1961sx,Bergmann:1968ve,Nordtvedt:1968qs,Wagoner:1970vr},  
written as effective Einstein equations, are   
\begin{eqnarray}
{\cal R}_{ab} - \frac{1}{2}\, g_{ab} {\cal R} &=& \frac{8\pi}{\phi} \,  
T_{ab}^\mathrm{(m)} + \frac{\omega}{\phi^2} \left( \nabla_a \phi 
\nabla_b \phi -\frac{1}{2} \, g_{ab} 
\nabla_c \phi \nabla^c \phi \right) \nonumber\\
&&\nonumber\\
&\, &  +\frac{1}{\phi} \left( \nabla_a \nabla_b \phi 
- g_{ab} \Box \phi \right) 
-\frac{V}{2\phi}\, 
g_{ab} \,, \nonumber \\
&& \label{BDfe1} 
\end{eqnarray}
\be
\Box \phi = \frac{1}{2\omega+3}   \left( 
8\pi T^\mathrm{(m)}   + \phi \, \frac{d V}{d\phi} 
-2V -\frac{d\omega}{d\phi} \, \nabla^c \phi \nabla_c \phi \right) \,, 
\label{BDfe2}
\ee
where ${\cal R}_{ab}$ is the Ricci tensor   
and $ T^\mathrm{(m)} $  is the 
trace of the matter stress-energy tensor 
$$T_{ab}^\mathrm{(m)}\equiv\frac{-2}{\sqrt{-g}}\, \frac{\delta 
S^\mathrm{(m)} }{\delta g^{ab} 
} \, .$$ The core idea of the thermal formulation is to group the 
contributions of the field $\phi$ and its derivatives into the effective 
stress-energy tensor 
\begin{eqnarray}
T_{ab}^{(\phi)} = \frac{\omega}{\phi^2} \left( \nabla_a \phi 
\nabla_b \phi - 
 \frac{1}{2} \, g_{ab} \nabla^c \phi \nabla_c \phi  \right) + 
 \frac{1}{\phi} \left( \nabla_a \nabla_b \phi -g_{ab} \square \phi \right) 
- \frac{V}{2 \phi} \, g_{ab} \,, \label{BDemt}
\end{eqnarray}
whose structure resembles the one of an imperfect fluid.

\section{Thermal view of scalar-tensor gravity \`a la Eckart}
\label{sec:2}
\setcounter{equation}{0}

Let us introduce the effective dissipative fluid corresponding to the 
Brans-Dicke-like scalar field $\phi$ which originates the thermal 
analogy, beginning with the kinematic quantities \cite{Ellis:1971pg}. 
Assuming that  the gradient $\nabla^a \phi$ is timelike and 
future-oriented, it defines the effective fluid 4-velocity  
\be u^{a}  = \frac{\nabla^a  \phi}{\sqrt{ -\nabla^e \phi \nabla_e \phi }} \,.
 \label{4-velocity}
 \ee
$\, $\\
Then spacetime splits into the time 
direction $u^c$ plus the  3-dimensional space seen by the observers 
comoving with the fluid and orthogonal to $u^a$.  The  
Riemannian metric in this 3-space is $h_{ab} \equiv g_{ab} + u_a u_b $, 
while ${h_a}^b$ is the projection operator onto it. The 
effective fluid 4-acceleration  $ \dot{u}^a \equiv u^b \nabla_b 
u^a $ is orthogonal to the 4-velocity. By projecting the 
velocity gradient  one obtains the tensor 
\be
V_{ab} \equiv  {h_a}^c \, {h_b}^d \, \nabla_d u_c 
= \Theta_{ab}= \sigma_{ab} +\frac{\Theta}{3} \, h_{ab}
\ee 
(where $\Theta \equiv {\Theta^c}_c =\nabla_c 
u^c$ is the expansion scalar), which 
coincides with the symmetric expansion tensor $\Theta_{ab}$. The vorticity 
$\omega_{ab} = V_{[ab]}$ vanishes because   
$u^a$ derives from a  gradient, then $u^a$ is irrotational and  
hypersurface-orthogonal \cite{Ellis:1971pg,Waldbook}.  $ \sigma_{ab} 
\equiv \Theta_{ab}-\Theta\, h_{ab}/3 $ is the  symmetric, trace-free shear 
tensor and the velocity gradient reads  
\cite{Ellis:1971pg}  
\be
\nabla_b u_a =  \sigma_{ab}+\frac{\Theta}{3} \, h_{ab} -  \dot{u}_a 
u_b  \,. 
\label{ecce}
\ee
Specializing these general definitions \cite{Ellis:1971pg,Waldbook} to 
the effective $\phi$-fluid, we compute 
\cite{Faraoni:2018qdr,Faraoni:2021lfc}
\begin{widetext}
\begin{eqnarray}
\dot{u}_a &=& \left( -\nabla^e \phi \nabla_e \phi \right)^{-2} 
\nabla^b \phi 
\Big[ (-\nabla^e \phi  \nabla_e \phi)  \nabla_a \nabla_b 
\phi + \nabla^c  \phi \nabla_b \nabla_c \phi \nabla_a \phi \Big] \,, 
\label{acceleration}\\
&&\nonumber\\
\Theta &=& 
\frac{ \square  \phi}{ \left (-\nabla^e \phi 
\nabla_e \phi \right)^{1/2} }  + \frac{ \nabla_a 
\nabla_b \phi \nabla^a \phi \nabla^b \phi }{ \left( -\nabla^e \phi 
\nabla_e 
\phi \right)^{3/2} } \,, \label{thetaScalar}\\
&&\nonumber\\
\sigma_{ab} &=&  \left( -\nabla^e \phi \nabla_e \phi \right)^{-3/2} \left[ 
-\left( \nabla^e  \phi \nabla_e 
\phi \right) \nabla_a \nabla_b  \phi 
- \frac{1}{3} \left(  \nabla_a \phi \nabla_b \phi   - g_{ab} \, \nabla^c 
\phi \nabla_c \phi   \right) \square \phi  \right.\nonumber\\
     &&\nonumber\\ 
     &\, & \left. - \frac{1}{3} \left( g_{ab} + \frac{ 2 \nabla_a \phi 
\nabla_b \phi }{   \nabla^e \phi \nabla_e \phi } 
 \right) \nabla_c \nabla_d 
\phi \nabla^d \phi \nabla^c \phi + \left( \nabla_a \phi \nabla_c 
\nabla_b 
\phi + \nabla_b \phi \nabla_c \nabla_a \phi \right) \nabla^c \phi \right]  
\,. \label{sheartensor}
\end{eqnarray} 

The Brans-Dicke field then generates the effective stress-energy 
tensor~(\ref{BDemt}) with imperfect fluid form 
\be 
T_{ab} = \rho u_a u_b + q_a u_b + q_b u_a + \Pi_{ab} 
\,, \quad\quad \quad\quad \Pi_{ab}= P  h_{ab} +\pi_{ab} 
\label{imperfectTab}
\ee 
and effective energy density, heat flux density, stress tensor, 
isotropic pressure, and anisotropic stresses 
\cite{Pimentel89,Faraoni:2018qdr,Giusti:2021sku} 
\begin{eqnarray}
 8\pi\,\rho^{(\phi)} &=&  -\frac{\omega}{2\phi^2} \, \nabla^e \phi \nabla_e 
\phi  +  \frac{V}{2\phi} + \frac{1}{\phi} \left( \square \phi -  
\frac{  \nabla^a \phi \nabla^b \phi \nabla_a 
\nabla_b \phi}{ \nabla^e \phi  \nabla_e \phi  } \right)  
\,,\label{effdensity}\\
&&\nonumber\\
 8\pi\,q_a^{(\phi)}   &=& \frac{\nabla^c  \phi \nabla^d \phi}{\phi 
  \left(-\nabla^e \phi \nabla_e \phi \right)^{3/2} } \,  
\Big(  \nabla_d \phi \nabla_c \nabla_a \phi 
- \nabla_a \phi \nabla_c \nabla_d \phi \Big) \,, \label{eq:q}\\
&&\nonumber\\
 8\pi\,\Pi_{ab}^{(\phi)}  &=&  
 \left( -\frac{\omega}{2\phi^2} \, \nabla^c \phi \nabla_c \phi 
-\frac{\Box\phi}{\phi} -\frac{V}{2\phi} \right) h_{ab} +\frac{1}{\phi} \, 
{h_a}^c {h_b}^d \nabla_c \nabla_d \phi \,,  \label{eq:effPi2}\\
&&\nonumber\\
 8\pi\,P^{(\phi)}  & = &  - \frac{\omega}{2\phi^2} \, \nabla^e \phi 
\nabla_e \phi - 
\frac{V}{2\phi} - \frac{1}{3\phi}  \left( 2\square \phi + 
\frac{\nabla^a \phi \nabla^b \phi \nabla_b \nabla_a \phi }{\nabla^e \phi 
\nabla_e  \phi }  \right) \,, \label{effpressure}\\
&&\nonumber\\
 8\pi\,\pi_{ab}^{(\phi)}   &=& \frac{1}{\phi \nabla^e \phi \nabla_e 
\phi } 
\left[  \frac{1}{3} \left( \nabla_a  \phi \nabla_b \phi - g_{ab} 
\nabla^c 
\phi \nabla_c \phi \right) \left(  \square \phi  - 
\frac{  \nabla^c \phi  \nabla^d \phi \nabla_d \nabla_c \phi }{ 
\nabla^e \phi 
\nabla_e \phi }   
\right) \right. \nonumber\\
&&\nonumber\\
&\, & \left. + \nabla^d \phi \left(  \nabla_d \phi \nabla_a \nabla_b 
\phi - 
\nabla_b \phi \nabla_a \nabla_d  \phi - \nabla_a \phi \nabla_d \nabla_b 
\phi +  
\frac{ \nabla_a \phi \nabla_b \phi  \nabla^c \phi \nabla_c 
\nabla_d \phi }{ \nabla^e \phi \nabla_e \phi } \right) \right] \,,
\label{piab-phi}
\end{eqnarray} 
\end{widetext}
while the isotropic pressure is the sum of perfect fluid and viscous 
contributions
\be
P^{(\phi)} = P^{(\phi)}_\mathrm{pf}  + P^{(\phi)}_\mathrm{v} \,.
\ee 
Let us briefly review  Eckart's thermal view of scalar-tensor 
gravity. 
Eckart's first order thermodynamics for {\em real} fluids  
(\cite{Eckart:1940te,Andersson:2006nr}) assumes three constitutive 
relations relating 
viscous pressure $P_\mathrm{v}$, heat current density $q^c$, and 
anisotropic 
stresses $\pi_{ab}$  to the expansion $\Theta$, temperature 
${\cal T}$, and shear 
tensor $\sigma_{ab}$: 
\begin{eqnarray}
P_\mathrm{v} &=& -\zeta \, \Theta \,,\label{def:Pv}\\
&&\nonumber\\
q_a &=& -{\cal K} h_{ab}\left(  \nabla^b {\cal T} + {\cal T} \dot{u}^b
\right) 
\,, \label{Eckart}\\
&&\nonumber\\
\pi_{ab} &=& - 2\eta \, \sigma_{ab} \,,\label{def:eta}
\end{eqnarray}
where $\zeta$ is the effective bulk viscosity coefficient, ${\cal K}$ is 
the effective thermal 
conductivity, and 
$\eta$ is the effective shear viscosity coefficient. $q^a$ is purely 
spatial in the frame of the effective fluid. The key point of the 
thermal analogy is that, by  comparing  
Eqs.~(\ref{eq:q}) and  (\ref{acceleration}), one miraculously obtains  
\cite{Faraoni:2018qdr} 
\be
q_a^{(\phi)} =  -\frac{ \sqrt{-\nabla^c \phi \nabla_c \phi}}{ 8 \pi \phi} \, 
\dot{u}_a \,.\label{q-a}  
\ee 
In the frame  comoving with the $\phi$-fluid,  the  spatial temperature 
gradient vanishes 
identically and the heat flow is only due to the inertia of energy.  
The temperature of the $\phi$-fluid, referred to as  the 
``effective temperature of 
scalar-tensor gravity'', pops out \cite{Faraoni:2018qdr} as
\be
{\cal KT}= \frac{ \sqrt{-\nabla^c \phi \nabla_c\phi}}{8\pi \phi} \,.
\label{temperature}
\ee
This quantity  is positive-definite and vanishes when the 
scalar $\phi$ is constant,  which reproduces  GR. Furthermore, the 
comparison of Eqs.~(\ref{piab-phi}) 
and~(\ref{sheartensor}) for  $\pi_{ab}^{(\phi)}$ and 
$\sigma_{ab}^{(\phi)}$ and the use Eq.~(\ref{def:eta}) give 
\be
\eta= - \frac{ \sqrt{-\nabla^c \phi \nabla_c \phi}}{16\pi \phi} 
= -\frac{ {\cal K}{\cal T}}{2} < 0 \,.
\ee
(Negative viscosities are common in fluid mechanics and  in non-isolated   
systems \cite{negviscosity}, and our $\phi$-fluid is not isolated due to 
the explicit coupling of $\phi$ to ${\cal R}$ in the action.)

The thermal view of scalar-tensor gravity \`a la 
Eckart includes an equation describing the 
approach to the GR equilibrium 
state, or its departure from GR 
\cite{Faraoni:2021lfc,Faraoni:2021jri,Giusti:2021sku}: 
\be
\frac{\d\left( K{\cal T}\right)}{\d\tau} = 8\pi \left( K{\cal T}\right)^2 
-\Theta\, {\cal K}{\cal T} +\frac{\Box\phi}{ 8\pi \phi} 
\,. \label{HeatEq}
\ee
In the simplified situations in which $\Box\phi=0$ one deduces 
\cite{Faraoni:2021lfc,Faraoni:2021jri,Giusti:2021sku} that near 
spacetime singularities, where the wordlines of the $\phi$ field 
converge ($\Theta<0$), the deviations of scalar-tensor gravity from GR are 
extreme. By contrast, the expansion of 3-space ($\Theta>0$) ``cools''  
gravity \cite{Faraoni:2021lfc,Faraoni:2021jri,Giusti:2021sku}.

\section{Thermal view of scalar-tensor gravity in the causal formalism}
\label{sec:3}
\setcounter{equation}{0}

A dissipative fluid in causal thermodynamics 
\cite{Maartens:1996vi,Andersson:2006nr} is 
described as follows.\footnote{This is the description of a {\em real}  
fluid and, at this stage, we do not yet apply it to the {\em effective} 
$\phi$-fluid of 
scalar-tensor gravity.} 
In causal thermodynamics, the Eckart constitutive relations \cite{Eckart:1940te} are promoted to the differential equations (e.g., 
\cite{Maartens:1996vi,Andersson:2006nr})
\begin{eqnarray}
 \tau_0 \, \dot{P}_\mathrm{v} +P_\mathrm{v} &=& -\zeta \Theta -\frac{\zeta 
{\cal T} }{2} \left[ 
\nabla_c \left( \frac{\tau_0 u^c}{\zeta {\cal T} } \right) \right] 
P_\mathrm{v} 
\,,\nonumber\\
&& \label{causthermo1}\\
 \tau_1 \, {h_a}^b \dot{q}_b + q_a &=& -{\cal K} \left( h_{ab} \nabla^b 
{\cal 
T} + {\cal T} \dot{u}_a \right) 
-\frac{ {\cal K}{\cal T}^2 }{2} \left[ 
\nabla_b \left( \frac{ \tau_1 u^b }{ {\cal K}{\cal T}^2 } \right)\right] 
q_a  \,,\label{causthermo2}\\
&&\nonumber\\
 \tau_2 \, {h_a}^c { h_b}^d \dot{\pi}_{cd} +\pi_{ab} &=& -2\eta \, 
\sigma_{ab} 
-\frac{\eta \, {\cal T} }{2} \left[ \nabla_c \left( 
\frac{ \tau_2 u^c}{ \eta {\cal T} } \right)\right] \pi_{ab} \,,\nonumber\\
&&\label{causthermo3}
\end{eqnarray}
where $\tau_0, \tau_1, \tau_2$ (which, in general, are not constant) are 
relaxation time scales and where an overdot denotes differentiation with 
respect to the proper time $\tau$ of the $\phi$-fluid, i.e., 
~$\dot{} \,\, 
\equiv u^c 
\nabla_c$.

In general, Eqs.~(\ref{causthermo1})--(\ref{causthermo3}) are difficult 
to solve analytically and  the purpose of this work is to find a special 
sector of the theory of causal thermodynamics that satisfies judicious, 
although restrictive, 
choices of constitutive relations. The result will then be applied to the 
effective $\phi$-fluid of scalar-tensor gravity. It is possible 
that  the thermal view ultimately makes sense for the effective fluid (our 
main subject of interest here) but only marginally for real fluids---this 
possibility will be assessed in future work.  Specifically, we look for 
the   special thermal scenarios satisfying 
\begin{eqnarray}
P_\mathrm{v} &=& -\zeta \Theta \,,\label{ansatz-Pv}\\
&&\nonumber\\
q_a &=& -{\cal K} \left( \nabla_a {\cal T} +{\cal T} \dot{u}_a \right) 
\,,\label{ansatz-q}\\
&&\nonumber\\
\pi_{ab} &=& -2\eta \sigma_{ab} \label{ansatz-pi} \,,
\end{eqnarray}
where now the heat flux density 4-vector is timelike, $q_c 
q^c <  0$. This causal vector field decomposes as 
\be
q_a = \left( -q_c u^c\right) u_a + {h_a}^b q_b \equiv 
q_a^{ (\parallel)} + q_a^{ (\perp)}  \label{q-decomposition}
\ee
with 
\be
 {h_a}^b q_b^{(\parallel)}=0 \,, \quad\quad  q_a^{(\perp)} u^a=0 \,.
\ee
The difference with the non-causal thermodynamics of Eckart \cite{Eckart:1940te} is that, while for Eckart $q^a$ is a purely spatial non-causal vector field satisfying the generalized Fourier law $ q_a = 
-{\cal K} h_{ab} \left(\nabla^b {\cal T} +{\cal T} \dot{u}^b \right)$ and implying superluminal heat flow 
\cite{Eckart:1940te}, now this vector 
field is promoted to a timelike one, with 
its spatial 
projection $q_a^{(\perp)}$ still satisfying the Eckart-Fourier law.  
By promoting $q^{a}$ to a time-like vector, equation~\eqref{ansatz-q} seems the most natural and straightforward generalization of the Eckart-Fourier
law and constitutes the most important and original assumption of this work. The scenario proposed here is a minimal 
generalization of Eckart's thermodynamics to make it causal. The fact that 
Eqs.~(\ref{ansatz-Pv}) and~\eqref{ansatz-pi} are still satisfied restricts 
the picture to Newtonian fluids, i.e., those in which viscous pressure and anisotropic stresses are {\it linear} in the gradient of the 
four-velocity, which is a statement about the mechanical behaviour of the 
fluid. Even in pre-relativistic physics, non-Newtonian fluids can be found 
among common materials (e.g., starch or custard) but Newtonian fluids are 
even more common and constitute the standard textbook situation (see e.g. \cite{Ruggeri:2024}). 
Eventually, we want to apply the causal generalization of Eckart's theory 
to the {\em effective} fluid of first-generation scalar-tensor gravity, 
which is well known to behave as a Newtonian fluid. More general theories, 
even in the well-explored ``viable'' Horndeski class, give rise to 
non-Newtonian 
effective (dissipative) fluids \cite{Miranda:2022wkz}.
  
Using the decomposition~(\ref{q-decomposition}), the dissipative fluid 
stress energy tensor 
\be
T_{ab} =\rho \, u_a u_b +Ph_{ab} +\pi_{ab} +q_a u_b +q_b u_a
\ee
with causal heat flux density $q_a$ retains the Eckart form and is  
rewritten as
\be
T_{ab} = \rho \, u_a u_b +P h_{ab} + \pi_{ab} + q_a^{(\perp)} 
u_b + 
q_b^{(\perp)} u_a \,, \label{newEckart}
\ee
which contains explicitly only the spatial part $q_a^{(\perp)} $ of $q_a$. 
The total energy density now reads
\be
\rho= \rho_\mathrm{pf} +\rho_\mathrm{v}  = \rho_\mathrm{pf} -2 q_c u^c 
\label{rhotot}
\ee
and, similar to the total pressure $P$, is the sum of the usual perfect 
fluid  energy 
density plus a ``viscous energy  density'' contribution $\rho_\mathrm{v}  
=-2q_cu^c $, which is non-negative since $q^a $ is now timelike and $q_a u^a <0$ to keep $q^a$ future-oriented. To 
this regard, the effective stress energy tensor $T_{ab}^{(\phi)}$ 
contains two terms: the perfect fluid contribution 
\be
 \frac{\omega}{\phi^2} \left( \nabla_a \phi 
\nabla_b \phi -\frac{1}{2} \, g_{ab} 
\nabla_c \phi \nabla^c \phi \right) -\frac{V}{2\phi}\, g_{ab}
\ee
similar to the stress-energy tensor of a minimally coupled matter 
scalar field, which is quadratic in the first order derivatives of 
$\phi$;    and the dissipative term
\be
 \frac{1}{\phi} \left( \nabla_a \nabla_b \phi 
- g_{ab} \Box \phi \right) 
\ee
linear in the second  derivatives of $\phi$, which is the only term 
describing dissipation. 
Accordingly, the effective fluid energy density and pressure 
$\rho^{(\phi)}$ and $P^{(\phi)}$ given by Eqs.~(\ref{effdensity}) and 
(\ref{effpressure}) clearly split into non-viscous and viscous 
contribution,  
while $q_a^{(\phi)}$ and $\pi_{ab}^{(\phi)} $ in 
Eqs.~(\ref{eq:q}) and (\ref{piab-phi}) contain second 
derivatives and are purely dissipative. 

Let us comment on the choices (\ref{ansatz-Pv})--(\ref{ansatz-pi}). 
Equations~(\ref{ansatz-Pv}) and~(\ref{ansatz-pi}) are 
Eckart's constitutive relations which mimic the 
three-dimensional non-relativistic 
relations of a Newtonian fluid and describe its mechanical properties, 
detailing the response to expansion, contraction, and shear 
deformations. The 
{\it ansatz}~(\ref{ansatz-q}) is the natural generalization of Eckart's 
law to 
the case of a timelike heat flux density (Eckart's law,  
in turn, generalizes Fourier's law of heat conduction). Since 
\be
q_a = - {\cal K} \left( \nabla_a {\cal T} + {\cal T}\dot{u}_a 
\right) = q_a^{(\parallel)} + q_a^{(\perp)}
\equiv  q^{(\parallel)} u_a  + q_a^{(\perp)} \,,
\ee
we have
\be
q^{(\parallel)} = -q_c u^c = {\cal K} \dot{\cal T} \,.
\ee

For scenarios satisfying Eqs.~(\ref{ansatz-Pv})--(\ref{ansatz-pi}), the 
causal thermodynamics equations~(\ref{causthermo1})--(\ref{causthermo3})   
reduce to 
\begin{eqnarray}
 \tau_0 \, \dot{P}_\mathrm{v} +  \frac{\zeta {\cal T} }{2} 
\left[  \nabla_c \left( \frac{\tau_0 \, u^c}{\zeta {\cal T} } \right) 
\right] P_\mathrm{v} =0 \,, \nonumber \\
&&\label{salcazzo1}\\
 \tau_1 \, {h_a}^b \dot{q}_b  + \frac{ {\cal K}{\cal T}^2 }{2} \left[ 
\nabla_b \left( \frac{ \tau_1 \, u^b }{ {\cal K}{\cal T}^2 } 
\right)\right]  q_a  -\left( q_c u^c \right) u_a  =0 \,, \nonumber\\
&&\label{salcazzo2}\\
 \tau_2 \, {h_a}^c { h_b}^d \dot{\pi}_{cd} 
+  \frac{\eta \, {\cal T} }{2} \left[ \nabla_c \left( 
\frac{ \tau_2 \, u^c}{ \eta {\cal T} } \right)\right] \pi_{ab} 
=0 \,.\nonumber\\
&&\label{salcazzo3}
\end{eqnarray}

Consider now the effective $\phi$-fluid of first-generation scalar-tensor 
gravity, for which the effective stress-energy tensor has the Eckart 
form~(\ref{newEckart}) 
\cite{Pimentel89,Faraoni:2018qdr,Giusti:2021sku}. By comparing 
the calculated heat flux density and 4-acceleration of $\phi$ 
\cite{Faraoni:2021lfc,Faraoni:2021jri,Giusti:2021sku}, 
\be
8\pi q_a^{(\phi)} = \frac{ \nabla^c \phi \nabla^d \phi}{\left( -\nabla^e 
\phi \nabla_e \phi \right)^{3/2} \phi} \left( \nabla_d \phi \nabla_a 
\nabla_c \phi - \nabla_a \phi \nabla_c \nabla_d \phi \right)
\ee
and  
\be
\dot{u}_a = \frac{ \nabla^b \phi \left[ \left( -\nabla^e\phi\nabla_e\phi 
\right) \nabla_a \nabla_b \phi + \nabla^c \phi \nabla_b \nabla_c \phi 
\nabla_a \phi \right]}{\left( -\nabla^e \phi \nabla_e \phi \right)^2} 
\ee
one obtains, similarly to non-causal thermodynamics\footnote{In what follows, we drop the $\phi$ superscript for the effective fluid to simplify the notation.}
\cite{Faraoni:2021lfc,Faraoni:2021jri,Giusti:2021sku},
\be
q_a^{(\perp)} =-{\cal K} {\cal T} \dot{u}_a
\ee
(with the difference that in non-causal thermodynamics 
$q_a$ coincides with its purely spatial projection $q_a^{(\perp)}$, while  
$q_a^{(\parallel)}=0$) and identifies again
\be
{\cal KT} =\frac{ \sqrt{ -\nabla^c \phi \nabla_c \phi}}{8\pi \phi} 
\label{KTagain}
\ee
for $\nabla^a \phi$ timelike and future-oriented. 
Furthermore, in the frame comoving with the $\phi$-fluid it is $h_{ab} 
\nabla^b {\cal T}=0$, that is, 
\be
{\cal T}={\cal T}\left( \tau \right) \,,
\ee
where $\tau$ is the proper time of the observer comoving with the 
$\phi$-fluid. This property does not imply that the choice 
(\ref{ansatz-Pv})--(\ref{ansatz-pi}) is limited to 
spatially homogeneous manifolds (e.g., FLRW 
or  Bianchi 
cosmologies) because the effective thermal conductivity ${\cal K}$ can 
still depend on the spatial position.  

Continuing, we have $q_a=q^{(\parallel)} u_a  +q_a^{(\perp)}$ with 
\begin{eqnarray}
q^{(\parallel)} &=& -q_c u^c ={\cal K} u^c \left( \nabla_c {\cal T} +{\cal T} \dot{u}_c \right) = {\cal K} \dot{\cal T} \,,\\[10pt]
q_a^{(\perp)} &=& {h_a}^b q_b = - {\cal K} {h_a}^b \left( \nabla_b {\cal T} +{\cal T} \dot{u}_b \right) =-{\cal K}{\cal T} \dot{u}_a \,,\\[10pt]
q_a &=& {\cal K} \dot{ {\cal T}} u_a -{\cal KT} \dot{u}_a  
\end{eqnarray}
and (given that $u^a$ is timelike and future-oriented) $q^a$ is future-oriented if $q^a u_a < 0$, i.e. ${\cal K} \dot{ {\cal 
T}}>0$, then
\be
\rho_\mathrm{v} = 2 q^{(\parallel)}= 2{\cal K} \dot{ {\cal T}} >0 \,.
\label{rhodef}
\ee
The relations between ${\cal KT}$ and the viscosity coefficients derived 
in \cite{Faraoni:2021lfc,Faraoni:2021jri,Giusti:2021sku}  for non-causal 
thermodynamics still hold in the new causal thermal 
view since they only involve purely spatial tensors:
\be
\zeta = -\frac{ {\cal KT}}{3} \,, \quad\quad 
\eta = -\frac{ {\cal KT}}{2} \,.
\ee
The differentiation of Eq.~(\ref{KTagain}) 
\cite{Faraoni:2021lfc,Faraoni:2021jri} along the effective fluid lines 
leads again to Eq.~(\ref{HeatEq}), which remains valid in the causal 
thermal view of scalar-tensor gravity.

\section{Bianchi~I cosmology}
\label{sec:4}
\setcounter{equation}{0}

Let us consider now spatially homogeneous but anisotropic Bianchi~I 
solutions of Brans-Dicke gravity,
\begin{eqnarray}
&& \d s^2=-\d t^2 +a_1^2(t) \d x^2 +a_2^2(t) \d y^2 +a_3^2(t) \d z^2 \,,\label{BianchiLineElement} \\[10pt]
&&\phi=\phi(t) \,. \label{PhiEquation}
\end{eqnarray}
Bianchi~I cosmologies are much simpler than a general geometry 
because they allow for shear and bulk viscosity but not for 
spatial heat fluxes. 

The scalar field gradient $\nabla^a \phi$ is timelike and is 
future-oriented if $\phi_t \equiv\partial_t \phi <0$. 
Since 
\be
u^0 = \frac{\d t}{\d\tau}= \frac{\nabla^0\phi}{ \sqrt{-\nabla^c \phi 
\nabla_c\phi}} =  \frac{g^{0a}\partial_a \phi}{\sqrt{\phi_t^2}} = 
\frac{-\phi_t}{| \phi_t |} 
= 1 \,,
\ee
one has $dt=d\tau$ and differentiation with respect to $t$ coincides with 
differentiation with respect to $\tau$ on scalar functions $f$,   
$ \dot{f}\equiv df/d\tau=df/dt$. We use the average scale 
factor 
\be
a(t) \equiv   \big[ a_1(t) a_2(t) a_3(t) \big]^{1/3} \,,
\ee
the average Hubble  function 
\be
H(t) \equiv \dot{a}/a 
\ee
 (in terms of which the 
expansion scalar is $\Theta=3H$), and the three directional Hubble   
functions 
\be
H_{(i)} \equiv \frac{ \dot{a}_{(i)} }{a_{(i)} } \quad\quad  (i=1,2,3).
\ee   

Assuming the relations~(\ref{ansatz-Pv})--(\ref{ansatz-pi}) and using the 
quantities
\be
X \equiv \frac{\tau_0}{3\zeta {\cal T}} \,,\quad \quad 
Y \equiv  \frac{\tau_1}{ {\cal K} {\cal T}^2} \,,\quad\quad 
Z  \equiv  \frac{\tau_2}{2\eta {\cal T}} \,,
\ee 
the reduced equations of causal 
thermodynamics~(\ref{salcazzo1})--(\ref{salcazzo3}) read, in a Bianchi~I 
universe, 
\begin{eqnarray}
&& \tau_0 \dot{P}_\mathrm{v} + \frac{\tau_0}{2X} 
\left[ \nabla_a \left( Xu^a \right) \right] P_\mathrm{v}=0 \,, 
\label{mondocane1}\\[10pt]
&& \tau_1 {h_a}^b \dot{q}_b + \frac{ \tau_1 }{2Y} 
\left[ \nabla_b \left( Y u^b \right) \right] q_a -\left( q_c u^c\right) 
u_a=0 \,,\label{mondocane2} \\[10pt]
&& {h_a}^c {h_b}^d \dot{\pi}_{cd} + \frac{1}{2Z} \left[ \nabla_e \left( Z 
u^e \right) \right] \pi_{ab}=0 \,,\label{mondocane3}
\end{eqnarray} 
which share a similar structure. 

Let us begin by analyzing Eq.~(\ref{mondocane1}), which is rewritten as 
\be
\frac{ \dot{P}_\mathrm{v} }{ P_\mathrm{v} } +\frac{3H}{2} 
+\frac{\dot{X}}{2X} =0 
\ee
and integrates to 
\be
|P_\mathrm{v}| = \frac{P_0}{a^{3/2} \sqrt{ |X|} } 
= \frac{P_0 \sqrt{ 3|\zeta| {\cal T} } }{ \sqrt{\tau_0} \, a^{3/2}}
\,,\label{BianchiPv}
\ee
where $P_0>0$ is an integration constant. Consistency with the assumption 
$P_\mathrm{v} = -\zeta \Theta = -3\zeta H $ then requires that 
\be
3|\zeta| |H| = \frac{P_0 \sqrt{ 3|\zeta| {\cal T} } }{ \sqrt{\tau_0} \, 
a^{3/2}} \, ,
\ee
where $\zeta =-{\cal KT}/3$, yielding the thermal conductivity
\be
{\cal K}= \frac{ K_0}{\tau_0 a \dot{a}^2}= \frac{ P_0^2}{\tau_0 a \dot{a}^2} \, ,
\,\label{BianchiK}
\ee
with ${\cal K}_0=P_0^2$. From this expression we derive the 
effective fluid temperature by remembering that ${\cal KT} = 
\sqrt{-\nabla^c \phi \nabla_c\phi}/\left( 8\pi \phi \right)$:
\be
{\cal T}=\frac{\tau_0}{8\pi P_0^2} \, a \dot{a}^2 \, 
\frac{ |\dot{\phi}|}{\phi}
\,. \label{BianchiT}
\ee
This result is particularly remarkable as it yields the first consistent 
splitting of ${\cal K}$ and ${\cal T}$, that is impossible to obtain (in a unique way) within the non-causal thermal formulation. Equations~(\ref{BianchiK}) 
and~(\ref{BianchiT}) for ${\cal K}$ and ${\cal T}$ can be substituted back 
into Eq.~(\ref{BianchiPv}) for $P_\mathrm{v}$, producing
\be
|P_\mathrm{v}| =  |H| \, \frac{ 
|\dot{\phi}|}{8\pi \, \phi}
\,. \label{BIviscouspressure}
\ee
Let us consider now Eq.~(\ref{mondocane2}), which we rewrite [using Eq.~\eqref{q-decomposition}] as 
\begin{eqnarray}
&& \tau_1 {h_a}^b \dot{q}_b + \frac{ \tau_1}{2Y} 
\left[ \nabla_c \left( Yu^c\right)\right] {h_a}^b q_b  + \left\{ \frac{\tau_1}{2Y} \left[ \nabla_c \left( Y 
u^c\right)\right]+1 
\right\} \left(-q_c u^c\right) u_a =0 \,.
\end{eqnarray}
A vector is identically zero if and only if all of its components vanish, hence this equality can be rewritten in terms of its projection in the direction of $u^a$ and onto the 3-space orthogonal to $u^a$, i.e.:
\begin{eqnarray}
&&{h_a}^b \tau_1 \dot{q}_b + 
\left\{ \frac{\tau_1}{2Y} \left[ \nabla_c \left( Yu^c\right)\right] q_b 
\right\} {h_a}^b =0 \,, \label{moh1}\\[10pt]
&& \left\{ \frac{\tau_1}{2Y} \left[ \nabla_c \left( Y u^c\right)\right] +1 
\right\}\left(- q_c u^c \right)=0 \,,\label{moh2}
\end{eqnarray}
respectively.
These equations admit the trivial solution $q_a=0$. 

Observe that, since $q_c u^c \neq 0$ we have that Eq.~(\ref{moh2}) implies
\be
\frac{\tau_1}{2Y}  \left[ \nabla_c \left( Y u^c\right)\right] =-1 \,.
\label{eq:inter}
\ee
Thus, plugging this condition into Eq.~\eqref{moh1} yields
\be
\left( \tau_1 \dot{q}_b -q_b\right) {h_a}^b=0 \,,
\ee
therefore the vector field $\tau_1 \dot{q}_b -q_b$ is parallel to the time direction, i.e. $\tau_1 \dot{q}_b -q_b = \lambda u_b$. However, for Bianchi I we have that $\dot{u}^a = u^c \nabla_c u^a = \nabla_0 u^a = 0$ since $u^a = \delta^a_{\,\,\, 0}$, hence the second term $ - 
q_b {h_a}^b = - q_a^{(\perp)}= -{\cal K} {\cal T} \dot{u}_a =0$. We therefore conclude that $\dot{q}_a $ is parallel to 
$u_a$.

Finally, let us discuss Eq.~(\ref{mondocane3}), which is rewritten as 
\be
{h_a}^c {h_b}^d \dot{\pi}_{cd} + \frac{1}{2} \left( \frac{ \dot{Z}}{Z}  
+3H \right) \pi_{ab}=0 \,,
\ee
or
\be
{h_a}^c {h_b}^d \dot{\pi}_{cd} 
+ \pi_{ab}  \left( \ln \sqrt{ |Z| a^3} \right)^{\bullet}  =0 \,.
\ee
It is shown in Appendix~\ref{Appendix:A} that $ \dot{\pi}_{ab}$ is always 
spatial, i.e., ${ h_a}^c {h_b}^d \dot{\pi}_{cd}= \dot{\pi}_{ab} $, turning 
this equation into 
\be
\dot{\pi}_{ab} 
+ \pi_{ab}  \left( \ln \sqrt{ |Z| a^3} \right)^{\bullet}  =0 
\,,\label{ToProve}
\ee
which admits a solution of the form
\be
\pi_{ab} = \frac{ C_{ab} }{\sqrt{ |Z| a^3} } \,,
\ee
where $C_{ab}$ is a constant symmetric and purely spatial 2-tensor. Using 
the definition of 
$Z$ and Eqs.~(\ref{BianchiK}) and (\ref{BianchiT}), this equation becomes
\be
\pi_{ab} = \bar{C}_{ab} \, \sqrt{ \frac{ \tau_0 }{\tau_2} } \, H \, 
\frac{ |\dot{\phi}|}{\phi} \, ,
\ee
absorbing some extra constants into the redefinition $\bar{C}_{ab}$ of $C_{ab}$ (and assuming $\tau_0, \tau_1, \tau_2 > 0$, which is reasonable since they are meant to represent time-scales).

\section{FLRW cosmology}
\label{sec:5}
\setcounter{equation}{0}

Let us consider now spatially flat 
Friedmann-Lema\^itre-Robertson-Walker universes, which are 
special cases of Bianchi~I metrics~(\ref{BianchiLineElement}) with 
$a_1(t)=a_2(t)=a_3(t) \equiv a(t)$.  Due to spatial homogeneity 
and isotropy, the shear $\sigma_{ab}$, anisotropic stresses 
$\pi_{ab}$, and spatial heat flux density $q_a^{(\perp)}$ all 
vanish, there is no shear viscosity and only bulk viscosity 
associated with the bulk viscous pressure $P_\mathrm{v}$ survives.

The expressions~(\ref{BIviscouspressure}) and~(\ref{BianchiT}) for 
the viscous pressure and temperature derived in Bianchi~I 
cosmologies still hold in the degenerate FLRW case. The effective 
pressure of the $\phi$-fluid, read from Eq.~(\ref{effpressure}), 
is clearly
\be
8\pi P_\mathrm{v} = -\frac{1}{3\phi} \left( 2\Box\phi +\frac{ 
\nabla^a\phi\nabla^b \phi \nabla_a\nabla_b \phi}{\nabla^e\phi 
\nabla_e\phi} \right) \,.
\ee
Let us restrict, for simplicity, to FLRW cosmologies in which 
$\Box \phi=-\left( \ddot{\phi}+3H\dot{\phi}\right)=0$, corresponding (cf. 
Eq.~(\ref{BDfe2})) to vacuum or conformal matter and zero or quadratic 
potential $V(\phi)$. In this case, the equation of motion $\Box\phi=0$  
gives the first integral $\dot{\phi}=\mbox{const.}/a^3$. In FLRW 
space we have 
\be
\nabla^a\phi\nabla^b \phi \nabla_a\nabla_b \phi = 
\dot{\phi}^2 \nabla_0\nabla_0 \phi = 
\dot{\phi}^2 \left( \ddot{\phi}- \Gamma^0_{00} \dot{\phi} \right)  
= \dot{\phi}^2 \ddot{\phi} \,,
\ee 
giving 
\be
P_\mathrm{v} = H \, \frac{ |\dot{\phi}|}{8\pi \phi} \,. 
\label{minchiadiSanGiuseppe}
\ee
As a check, this time dependence matches that of 
Eq.~(\ref{BIviscouspressure}). 

The heat flux density $q^a=\left( q^0, \bm{0}  \right)$  has only 
the time component while for $u^a$ to be future-oriented we need $\dot{\phi} < 0$, thus 
\begin{eqnarray}
q^{(\parallel)} 
&=& {\cal K}\dot{\cal T} = \frac{P_0^2}{\tau_0\, a \dot a^2}
\frac{\d}{\d t}\left(
-\frac{\tau_0}{8\pi P_0^2}\, a \dot a^2 \frac{\dot\phi}{\phi}
\right)
= -\frac{1}{8\pi}\,
\frac{1}{\tau_0\, a \dot a^2}
\frac{\d}{\d t}\left(
\tau_0\, a \dot a^2 \frac{\dot\phi}{\phi}
\right) \nonumber
\\[6pt]
&=& -\frac{1}{8\pi}\,
\frac{1}{\tau_0\, a \dot a^2}
\Bigg[
\dot\tau_0\, a \dot a^2 \frac{\dot\phi}{\phi}
+ \tau_0 \frac{\d}{\d t}(a\dot a^2)\frac{\dot\phi}{\phi}
+ \tau_0 a\dot a^2 \left(\frac{\dot\phi}{\phi}\right)^{\bullet}
\Bigg] \nonumber
\\[6pt]
&=& -\frac{1}{8\pi}\Bigg[
\frac{\dot\tau_0}{\tau_0}\frac{\dot\phi}{\phi}
+ \frac{1}{a\dot a^2}\frac{\d}{\d t}(a\dot a^2)\frac{\dot\phi}{\phi}
+ \left(\frac{\dot\phi}{\phi}\right)^{\bullet}
\Bigg] \nonumber \\[6pt]
&=&
-\frac{1}{8\pi}\frac{\dot\phi}{\phi} 
\Bigg[
\frac{\dot\tau_0}{\tau_0}
+ \frac{1}{a\dot a^2}\frac{\d}{\d t}(a\dot a^2)
+ \frac{\phi}{\dot\phi}\left(\frac{\dot\phi}{\phi}\right)^{\bullet}
\Bigg] \nonumber \\ [6pt]
&=&
-\frac{\dot\phi}{8\pi \phi}
\,
\frac{\d}{\d t}
\ln\!\left|\frac{a\dot a^2 \tau_0 \, \dot{\phi}}{\phi}\right|.
\end{eqnarray}
If we now use the assumption $\Box \phi = 0$ we find that $\dot{\phi} = v_0/a^3$, for which we conclude that
\begin{eqnarray}
q^{(\parallel)}
= -\frac{\dot\phi}{8\pi \phi}
\,
\frac{\d}{\d t}
\ln\!\left|\frac{a\dot a^2 \tau_0 \, \dot{\phi}}{\phi}\right| =
-\frac{\dot\phi}{8\pi \phi}
\,
\frac{\d}{\d t}
\ln\!\left|\frac{H^2 \tau_0}{\phi}\right| \, . \label{qpall-dupall}
\end{eqnarray}
Let us consider now the viscous component of the 
effective  energy density~(\ref{effdensity}):
\be
\rho_\mathrm{v} = \frac{1}{8\pi \phi} \left( \Box\phi - \frac{ 
\nabla^a \phi \nabla^b \phi \nabla_a\nabla_b \phi}{\nabla^e\phi 
\nabla_e\phi} \right) = -\frac{3H\dot{\phi}}{8\pi \phi} = 3P_\mathrm{v} \, .
\label{minchiadiBabboNatale1}
\ee
We also have 
\begin{eqnarray}
\rho_\mathrm{v} &=& 2{\cal K} \dot{{\cal T}} =
-\frac{\dot{\phi}}{4\pi\phi} \, \frac{\d}{\d t}\ln \left| \frac{H^{2}\tau_0}{\phi} \right| 
\,.\label{minchiadiBabboNatale2}
\end{eqnarray}
For consistency, the two expressions of $\rho_\mathrm{v}$ 
(\ref{minchiadiBabboNatale1}) and 
(\ref{minchiadiBabboNatale2}) must be equal, which provides an 
equation for the relaxation time scale $\tau_0$:
\be
-\frac{3H\dot{\phi}}{ 8\pi \phi} = -\frac{ \dot{\phi}}{4\pi\phi} 
\, \frac{\d}{\d t} \ln \left| \frac{H^{2}\tau_0}{\phi} \right| 
\,,
\ee  
or
\be
\frac{ 3 \,\dot{a}}{2 \,a} ={\frac{\d}{\d t} \ln \left| \frac{H^{2}\tau_0}{\phi} \right| } \,,
\ee  
which implies
$$
\frac{\d}{\d t} \ln a^{3/2} = \frac{\d}{\d t} \ln \left| \frac{H^{2}\tau_0}{\phi} \right|  \, ,
$$
which yields
\be
\tau_0 = \frac{C_{0} \, a^{3/2} \, \phi}{{H}^2}=\frac{C_{0} \, a^{7/2}\,\phi}{\dot{a}^{2}}
\ee
with $C_{0}>0$ an integration constant. Moreover the logarithmic derivative of $\tau_{0}$ yields
\begin{align}
\frac{\dot {\tau}_{0}}{\tau_{0}}&=\frac{3H}{2} -\frac{2\dot H}{H}-\frac{|\dot \phi|}{\phi} =\frac{\Theta}{2}-\frac{2\dot \Theta}{\Theta}-8\pi{\cal KT} \, .
\label{tauev}
\end{align}
This equation highlights a competition between the first  two terms and 
the third one. In an expanding universe, the expansion terms consistently 
provide a positive a contribution, whereas the third term  always 
contributes negatively. This interplay serve as a benchmark  for 
understanding when the effective Eckart theory properly  approximates the 
Israel-Stewart generalization introduced here. Indeed, during the  relaxation process, 
when GR is approached, ${\cal KT} \to 0$. In this limit  the expansion 
terms becomes dominant, leading to $\dot{ \tau}_{0}>0$,  which makes the 
causal formulation relevant. Conversely, in strong gravity  regimes far 
from GR (${\cal KT} \to  \infty$) the negative term dominates  leading to 
$\dot{ \tau}_{0}<0$. In this scenario the Eckart theory provides a good approximation of the causal thermal setup.
\par 
Finally, let us divide 
Eqs.~(\ref{minchiadiSanGiuseppe}) and 
(\ref{minchiadiBabboNatale1}) to obtain 
\be
\frac{P_\mathrm{v}}{\rho_\mathrm{v}} =\frac{1}{3} \,,
\ee
that is, the viscous component of the effective $\phi$-fluid  
obeys the radiation equation of state. 
One 
cannot ascribe much physical meaning to this feature because 
$P_\mathrm{v}$ and $\rho_\mathrm{v}$ always appear together with 
their non-viscous counterparts and their splitting into 
non-viscous and viscous parts is formal. The exception is the 
case $\omega=0$, $V=0$ in which only the viscous components are 
present (which could be called a ``maximally viscous'' scenario). $f(R)$ 
gravity is equivalent to an $\omega=0$ Brans-Dicke 
theory, but  with a complicated potential $V(\phi)$  
\cite{Sotiriou:2008rp,DeFelice:2010aj,Nojiri:2010wj} which reintroduces 
a non-viscous term. The non-causal thermal view of $f(R)$ cosmology is 
discussed in \cite{Faraoni:2025fjq}. We leave the causal thermal analysis 
of $f(R)$ gravity for future work.

\section{Exact solutions}
\label{sec:6}

In this final section, we consider exact solutions that provide an 
explicit expression for $\phi$ and allow for the analytical computation of 
the thermodynamic quantities. 
\vfill\null \noindent {\em Kasner solution.}---The Kasner vacuum solution 
in GR is a special case of the spatially homogeneous and anisotropic 
Bianchi~I geometry, which generalizes the FLRW spatially flat universe. 
The line element in comoving Cartesian coordinates reads 
\be
\d s^{2}=-\d t^{2}+a_{1}^{2}(t)\d x^{2}+a_{2}^{2}(t)\d 
y^{2}+a_{3}^{2}(t)\d z^{2}, 
\ee
with
\be
a_{i}=\left(\frac{t}{t_{0}}\right)^{p_{i}} \,,  \quad i=1,2,3
\ee
and where $t_{0}$ and $p_{i}$ are constants obeying
\be
\label{pis}
p_{1}+p_{2}+p_{3}=1, \quad \quad p_{1}^{2}+p_{2}^{2}+p_{3}^{2}=1 \,.
\ee 
However, in scalar-tensor gravity with constant $\omega$, $V(\phi)=0$ and 
$\phi=\phi(t)$ (because of spatial homogeneity), one obtains 
instead (\cite{Ruban:1974eb,Ruban:1972bg}) 
\be
a_{i}=\left(\frac{t}{t_{0}}\right)^{\frac{p_{i}}{1+C}}\,, \quad  \quad 
\phi(t)=\phi_{0}\left(\frac{t}{t_{0}}\right)^{\frac{C}{1+C}}
\ee
with $\phi_{0}$ and $C$ constants.\footnote{Setting $C=0$ reduces the 
solution to GR.} This gives the average scale factor 
\be
a(t)=\left(\frac{t}{t_{0}}\right)^{\frac{1}{3(1+C)}}\, ,
\ee
where Eq.~$\eqref{pis}$ has been used.
\par
The future-orientedness of $\nabla^a\phi$ requires $\dot \phi<0$, 
namely
\be
\dot{\phi}=\frac{C}{1+C}\frac{\phi_{0}}{t_{0}} 
\left(\frac{t}{t_{0}}\right)^{-\frac{1}{1+C}}<0 \, ,
\ee 
that it is satisfied for
\be
-1<C<0 \,.\label{restriction}
\ee
With these prescriptions, one can now compute the (effective) 
dissipative quantities $\rho_{\mathrm{v}},  P_{\mathrm{v}}, {\cal 
K},{\cal T}, q^{(\parallel)}$ and ${\cal KT}$. However, since ${\cal K}$ 
and ${\cal T}$ depend on $\tau_{0}$, we first compute the latter
\begin{eqnarray}
\tau_{0} &=& 9 (C+1)^2 C_{0} t_{0}^{2} \phi_{0} \left(\frac{t}{t_{0}}\right)^{\frac{6C+5}{2 (C+1)}} \,,\\[10pt]
\rho_{\mathrm{v}} &=& 3P_{\mathrm{v}}=\frac{|C| \,}{8 \pi (C+1)^2  \, t^2} \, ,\\[10pt]
{\cal K} &=& \frac{P_{0}^{2}}{\phi_{0} C_{0}} \left(\frac{t}{t_{0}}\right)^{-\frac{3+2C}{2(1 + C)}} \, ,\\[10pt]
{\cal T} &=& \frac{|C| \, C_{0} \, \phi_{0}}{8 \pi (C+1) P_{0}^{2} t_{0}}\left(\frac{t}{t_{0}}\right)^{\frac{1}{2(1+C)}}  \, , \\[10pt]
\K\T &=& \frac{|C|}{8\pi \, t \, (1+C)} \,.
\end{eqnarray}
The restriction~(\ref{restriction}) on $C$ ensures the positivity of all 
these quantities. At late times, both the viscous components and ${\cal 
KT}$ relax to zero, recovering GR. This situation is consistent with a 
perfect fluid reaching a thermal stable equilibrium. From a thermodynamic 
perspective, this limit corresponds to the relaxation of a dissipative 
process driven by the cosmic expansion. Conversely for $t\to 0^{+}$, all the thermodynamic quantities diverge, with the exception of $\T$ that vanishes. Moreover, $\tau_{0}$ does not vanish in the $t\to +\infty$ limit for $C\in[-5/6,0)$, namely in the parameter space closest to GR. This agrees with Eq.~\eqref{tauev}, indicating that the causal model becomes increasingly relevant as the theory approaches GR. Finally, the time component of 
the heat flux reads
\be
q^{(\parallel)}=\frac{|C|}{16\pi (C+1)^{2}t^{2}} \, , 
\ee
that matches $3P_{\mathrm{v}}=\rho_{\mathrm{v}} = 2q^{(\parallel)}$. 
\vfill\null
\noindent {\em O'Hanlon and Tupper (O'H-T) solution.}---The O'H-T vacuum solution 
\cite{OHanlon:1972ysn} represents a flat FLRW power-law universe of 
Brans-Dicke cosmology obtained for $\omega=\text{const.}>-3/2$, $V=0$, and
\be
a=\left(\frac{t}{t_{0}}\right)^{q_{\pm}}, \quad \phi=\phi_{0} 
\left(\frac{t}{t_{0}}\right)^{s_{\pm}} \,, 
\ee
with
\be
s_{\pm}=1-3q_{\pm}, \quad q_{\pm}=\frac{\omega}{3(\omega+1)\mp 
\sqrt{3(2\omega+3)}} \, .
\ee
The request that $\nabla^{a}\phi$ be future-oriented forces the selection 
of the $(+)$ solution, yielding 
\begin{eqnarray}
\rho_{\mathrm{v}} &=& 3P_{\mathrm{v}}=\frac{\,|s\,(s-1)|}{8\pi 
t^{2}}\, , \\[10pt]
\tau_{0} &=& \frac{9 C_{0}\, \phi_{0}\, t_0^{2}}{(s-1)^{2}}\left(\frac{t}{t_{0}}\right)^{\frac{5+s}{2}}\, ,\\[10pt]
{\cal K} &=& \frac{P_{0}^{2} }{C_{0} \phi_{0} 
}\left(\frac{t}{t_{0}}\right)^{\frac{s-3}{2}}\, ,\\[10pt]
{\cal T} &=& \frac{C_{0} \phi_{0} |s| }{8 \pi  
P_{0}^{2}t_{0}}\left(\frac{t}{t_{0}}\right)^{\frac{1-s}{2}}\, ,\\[10pt]
\K\T &=& \frac{|s|}{8\pi t}\, .
\end{eqnarray}
Both $\rho_{\mathrm{v}}$ and $P_{\mathrm{v}}$ vanish for $s \to 0$ 
(i.e., $\phi \to \phi_{0}$) and for $t \to \infty$,  stressing their 
dissipative nature. Notice that $s-1=0$ would require  $q=0$, which is 
impossible  for $\omega \neq 0$. The decoupling of ${\cal K}$ and ${\cal 
T}$ shows that, at any given finite time $t$, ${\cal T}$ is the only 
parameter that provides relaxation toward GR in the $s\to 0$ 
limit and again exhibits a regular behavior as $t\to0^{+}$. 
{\em Vice-versa}, ${\cal K}$ vanishes as $t\to +\infty$ and diverges as $t\to 0^{+}$. For this solution $\tau_0$ never vanishes at late times, whereas it does in the strong gravity regime as $t\to0^{+}$ when Eckart theory well approximates Israel-Stewart. Finally, the time component of $q^{a}$, 
\be
q^{(\parallel)}=\frac{ |s (s-1)|}{16 \pi t^{2}}\, ,
\ee
shares the same analysis with $\rho_{\mathrm{v}}$ and $P_{\mathrm{v}}$. 
\section{Conclusions}
We have extended the thermal analogy for the effective $\phi$-fluid from 
Eckart's non-causal formulation to the Israel-Stewart causal 
framework, adopting the minimal {\it ansatz} of promoting the heat 
flux 
density from a purely spatial vector to a timelike one, 
$q_a=(q^{(\parallel)},q_a^{(\perp)})$. By restricting our analysis to a 
subclass of theories where the dissipative components still obey Eckart's 
constitutive relations \eqref{def:Pv}-\eqref{def:eta}, the spatial 
projection $q^{(\perp)}_a$ preserves the definition of the effective 
temperature \eqref{temperature}. Thus, the product ${\cal KT}$ maintains 
its crucial role as the order parameter governing the relaxation toward 
General Relativity. Meanwhile, the time component $q^{(\parallel)}$ enters 
the Israel-Stewart equations and provides a clear physical interpretation of the 
``viscous'' effective density, $\rho_\mathrm{v}=2q^{(\parallel)}$. 
Specializing this setup to Bianchi~I and FLRW cosmologies allowed us to 
uniquely decouple $\cal{K}$ and $\cal{T}$ for the first time. This 
approach also naturally yields an expression for the relaxation time 
$\tau_0$, whose evolution defines the regimes where Eckart's theory 
remains a valid approximation of the Israel-Stewart model.

A precise mapping of scalar-tensor gravity onto the Israel-Stewart constitutive 
equations would require higher-order derivatives of $\phi$. While these 
could be introduced via higher-order terms in the Lagrangian without 
suffering from the Ostrogradsky instability, doing so would 
fundamentally disrupt the standard effective dissipative fluid structure. 
Therefore, our {\it ansatz} constitutes the minimal viable step 
toward a consistent causal formalism. Due to the purely formal nature of 
this analogy, we do not attempt to draw connections to a physical entropy, 
as such a quantity lacks a rigorous thermodynamic meaning in this {\em 
effective} fluid context. Future works will apply the new causal thermal 
framework to $f(R)$ gravity and will further investigate its implications 
for cosmological singularities. 

A relevant question is the extent to which the {\em ansatz}
 (\ref{ansatz-Pv})-(\ref{ansatz-pi})  applied to the {\em effective} 
fluid of scalar-tensor gravity, which we ended up working with, is 
applicable to {\em real} fluids. In the end, mathematical statements about 
the Israel-Stewart (or any other causal) formalism should be corroborated 
by rigorous mathematical theorems and by kinetic theory. The latter is not 
available for the {\em effective} $\phi$-fluid of scalar-tensor gravity 
because it is not made of particles like a real fluid. The {\it ansatz} 
(\ref{ansatz-Pv})-(\ref{ansatz-pi})  works for the effective 
fluid, leading to exact solutions, but at this stage one cannot be sure 
about real fluids. This question will be pursued
 in the future.

\section*{Conflict of interest}

The authors declare no conflict of interest.

\section*{Data Availability Statement}

Data sharing is not applicable to this article as no new data were created or analyzed in this study. 

\section*{acknowledgments} 

The work of L.G. and A.G. has been carried out in the framework of activities of 
the National Group of Mathematical Physics (GNFM, INdAM). A.G. is supported by the Italian Ministry of Universities and Research (MUR) through the grant ``BACHQ: Black Holes and The Quantum'' (grant no. J33C24003220006) and by the INFN grant FLAG. V.F. is supported by the Natural Sciences \& Engineering Research Council of Canada (Grant no. 2023--03234).

\medskip

\begin{appendices}

\section{Proof of Eq.~(\ref{ToProve}) }
\label{Appendix:A} 
\renewcommand{\theequation}{A.\arabic{equation}} 
\setcounter{equation}{0}

First, recall that we are working in Bianchi I and that $u^a = \delta^a _{\,\,\, 0}$. Thus we have that
\begin{eqnarray}
\dot{\pi}_{ab} & \equiv & u^c \nabla_c \pi_{ab} = u^c \left( \partial_c 
\pi_{ab}
-\Gamma_{\,\,ca}^d \pi_{db} - \Gamma^d_{\,\,cb} \pi_{ad} \right) \nonumber\\
&=& \partial_0 
\pi_{ab}
-\Gamma_{\,\,0a}^d \pi_{db} - \Gamma^d_{\,\,0b} \pi_{ad} \,,
\end{eqnarray}
The only non-vanishing Christoffel symbols of the Bianchi~I metric~\eqref{BianchiLineElement} read
$$
\Gamma ^i _{\,\,0j} = H_{(i)} \, \delta^i _{\,\, j} \, ,
$$
for $i,j = 1,2,3$. This implies that
$$
\dot{\pi}_{00} = 0 \quad \mbox{and} \quad \dot{\pi}_{0j} = 0 \, ,
$$
again with $j = 1,2,3$. Whereas, it is easy to see that
$$
\dot{\pi}_{ij} = \partial_0 \pi_{ij} - \big[H_{(i)} + H_{(j)}\big] \, \pi_{ij} \, ,
$$
with $i,j = 1,2,3$. Therefore, $\dot{\pi}_{ab}$ has no non-vanishing time-time or 
time-space components and ${h_a}^c {h_b}^d \dot{\pi}_{cd} = 
\dot{\pi}_{ab}$.
\newline
\end{appendices}

\end{document}